\documentclass[aps,preprint,letterpaper,amssymb,showkeys,10pt]{revtex4}

\usepackage{amsmath}

\begin{document}
\title{Generalized Number Theoretic Spin Chain-Connections to Dynamical Systems and Expectation Values}
\author{Jan Fiala}
\affiliation{Department of Physics \& Astronomy, University of Maine, Orono, ME 04469.  Current address: Department of Physics, Clark University}
\email{jfiala@clarku.edu; phone: (508) 793-7759}
\author{Peter Kleban}
\affiliation{LASST and Department of Physics \& Astronomy, University of Maine, Orono, ME 04469}
\email{kleban@maine.edu; fax:  (207) 581-2255; phone:  (207) 581-2258}

\date{\today}
\begin{abstract} 
We generalize the number theoretic spin chain, a one-dimensional statistical model based on the Farey fractions, by introducing a new parameter $x \ge 0$.  This allows us to write recursion relations in the length of the chain. These relations are closely related to the Lewis three-term equation, which is useful in the study of the Selberg $\zeta-$function.  We then make use of these relations and spin orientation transformations.  We find a simple connection with the transfer operator of a model of intermittency in dynamical systems.  In addition, we are able to calculate certain spin expectation values explicitly in terms of the free energy or correlation length.  Some of these expectation values appear to be directly connected with the mechanism of the phase transition.
 \end{abstract}

\maketitle

\section{Introduction}\label{treti}

In this paper, following a suggestion of Zagier \cite{Don}, we generalize the ``number theoretic" partition function, whose statistical mechanical properties have been studied by Knauf \cite{K,C-K,K-o1,C-Kn}, by introducing a new parameter $x\in \mathbb R^+_0$. Both the ``canonical" and ``grand canonical" partition functions of Knauf arise, for certain values of $x$.
More generally, as explained in  section \ref{pf}, these models may be regarded as one-dimensional spin chains of length $k$, and the new parameter $x$ allows us to derive recursion relations on the length  of the spin chain.  These relations are simple generalizations of the Lewis three-term equation that has been extensively studied in number theory \cite{L-Z, LZ}.  Next, section \ref{2}   explores some simple consequences of spin orientation (``spin flip") transformations for partition functions and expectation values. These results, along with the recursion relations, are our main tools.  They are used, in various ways, in the succeeding sections.  In  section \ref{3}, 
the recursion relations are shown to imply a very simple and direct connection between the transfer operator studied by Prellberg \cite{P-S, P-s, Diss} in a model of dynamical systems with intermittency and the generalized partition function.  This is one of our main results.  We examine some of its consequences.  In particular, we prove that all models have the same free energy (and hence the same thermodynamics), independent of the value of $x$, and show that the known spectrum of the transfer operator implies that the correlation length satisfies the prediction of scaling theory. In sections \ref{4} and \ref{5}, we use the recursion relations and spin-flip behavior to calculate certain spin expectation values  for both finite and infinite spin chains.  Some new and interesting features arise.  In particular, the expectation value of certain spin clusters and an independent spin at arbitrary distance is shown to be independent of the direction of the spin, at all temperatures above the transition. Thus, in this sense, the spin cluster removes the spin asymmetry of the system.  This behavior appears to be related to the mechanism of the phase transition.

All of our results are rigorous.  Since they concern certain weighted averages over Farey fractions and relate to the Lewis equation, they may be of interest to mathematicians.  Therefore we have included a few explanations and definitions in an attempt to make the paper more accessible to those unfamiliar with statistical mechanics.  
 
\section{Definition of the partition function}\label{pf}
In this section we define the generalized partition function, and show that it satisfies a recursion relation.  This relation is one of our main tools for proving new results. 

Let the matrix $M_k$ be any product of $k$ matrices
$A_0:=\left({1\atop 1}{0\atop 1}\right)$ and $A_1:=\left({1\atop 0}{1\atop 1}\right)$,
$$M_k=\left ( \begin{array}{cc}
a&b\\
c&d
\end{array}
\right ).$$
We can regard any such product as a one-dimensional chain of length $k$.  If the $i$th matrix ($1 \le i \le k$) is $A_0$  we identify it as a spin pointing up at the $i$th site in the chain, and likewise as a spin pointing down if it is $A_1$. Therefore each matrix $M_k$ corresponds to a definite configuration of the $k$ spins.

Next we extend the Knauf model \citep{K-o1}   (see also \citep{K-O} and  (\ref{KnaufP})) 
by introducing a family of partition functions parametrized by the variable $x \ge 0$ \cite{Don}
\begin{equation}\label{def}
\tilde{Z}_k(x,\beta):=\sum (cx+d)^{-2\beta},
\end{equation}
where the sum runs over all $2^k$ permutations of the product of the $k$ matrices $A_0$, $A_1$.

Next,  setting $M_{k+1}=M_kA_0$ or $M_{k+1}=M_kA_1$ gives rise to a chain of  length $k+1$.  Now since 
$M_kA_0=
\left({a+b\atop c+d}{b\atop d}\right)$ 
and
$M_kA_1=
\left({a \atop c}{a+b\atop c+d}\right)$,
 we find from (\ref{def}) the recursion relation
\begin{equation}\label{rec}\tilde{Z}_{k+1}(x,\beta)=
(1+x)^{-2\beta}\tilde{Z}_{k}\left(\frac{x}{1+x},\beta \right)+
\tilde{Z}_{k}(x+1,\beta)
\end{equation}
with the initial condition $\tilde{Z}_{0}(x,\beta)\equiv 1$ (i.e.~$M_0=\left({1\atop 0}{0\atop 1}\right)$).
The variable $x \in \mathbb R^+_0$ is a parameter which changes the energy ($E_k=2 \ln(cx+d)$)  of each spin configuration (i.e.~each matrix product $M_k$)
and $\beta \in \mathbb R^+_0$ is the inverse temperature.   However, the thermodynamics is independent of $x$, as we will see.

It is convenient to define, as in number theory \cite{NT}, the action of the matrix $M=\left({a\atop c}{b\atop d}\right)$ 
on any function $f(x)$
\begin{equation}\label{action}f(x)|M:=(cx+d)^{-2\beta}f \left (\frac{ax+b}{cx+d}\right ).
\end{equation}
For example, consider the action of the matrix $A_0$ on the constant function 
\begin{equation}\label{initialcon}1(x)|A_0=(1+x)^{-2\beta},
\end{equation} 
where $1(x)\equiv 1$.

It is easy to check that our partition function $\tilde{Z}_k(x,\beta)$ can be written as
\begin{equation}\label{pfac}\tilde{Z}_k(x,\beta)=\sum_{i=1}^{2^k}1(x)|M_i,\end{equation}
where $M_i=\Pi_{j=1}^{k}A_{\tau_j(i)}$ with $\tau_j(i)\in \{0,1\}$. Note that each $M_i$ defines fractions $\frac{a}{b}$ and $\frac{c}{d}$ at level $k$ of the Stern-Brocot tree \cite{CM}; thus the level corresponds to the length of the spin chain.  The subset of these fractions between zero and one are called Farey fractions. They are generated by the products which start with $A_0$ \citep{K-O}.

In the following, we make extensive use of an abbreviated form of (\ref{pfac})
\begin{equation}\label{abpfac}
\tilde{Z}_k(x,\beta)=1(x)|(A_0+A_1)^k=
1(x)|A_0(A_0+A_1)^{k-1}+
1(x)|A_1(A_0+A_1)^{k-1},
\end{equation}
where the addition must be applied {\bf after} the multiplication of the matrices!

The Knauf ``canonical" partition function $Z^K_k(s)$ (see \citep{K-o1} for the definition and note that $s=2\beta$) is equal to 
\begin{equation}\label{KnaufP}
Z_k^K(2\beta)=
1(x)|A_0(A_0+A_1)^{k}|_{x=0}.
\end{equation}
Similarly, the ``grand canonical" partition function of Knauf \citep{K}
corresponds to (\ref{KnaufP}) with $x=1$ on the right hand side. 

Let 
\begin{equation}\label{defofpartfunction}
Z_k(x,\beta):=1(x)|A_0(A_0+A_1)^{k}.
\end{equation}
(\ref{defofpartfunction}) is then a direct generalization of $Z_k^K$.  It is easy to verify that (\ref{defofpartfunction}) satisfies the recursion relation (\ref{rec}).
Using $1(x)|A_1=1(x)$ and (\ref{abpfac}) we get 
\begin{equation}\label{fromgKnauf}
\tilde{Z}_k(x,\beta)=
Z_{k-1}(x,\beta)+
1(x)|A_1(A_0+A_1)^{k-1}=
1+\sum_{i=0}^{k-1}Z_{i}(x,\beta).
\end{equation}
Thus (\ref{fromgKnauf}) relates two partition functions satisfying the recursion formula (\ref{rec}) with initial conditions
$Z_0(x,\beta)=(1+x)^{-2\beta}$ and $\tilde{Z}_0(x,\beta)=1(x)$. 

In the following, we use $Z_k(x,\beta)$ exclusively, because of its direct relation to previously studied spin chains.  It is possible to obtain similar results for $\tilde{Z}_k(x,\beta)$  as well. We show in section \ref{3} that all of the partition functions (\ref{defofpartfunction}) have the same free energy, i.e.~their thermodynamics is independent of $x$ (for the definition of free energy, cf.~(\ref{freeEF})).

For use below, we note that it is straightforward to express $Z_k(x,\beta)$  similarly to  $\tilde Z_k(x,\beta)$ in (\ref{def}) by
\begin{equation}\label{def2}
{Z}_k(x,\beta)=\sum ((a+c)x+(b+d))^{-2\beta},
\end{equation}
where ${a,b,c,d}$ now correspond to the matrix elements of $M_{k-1}$. Note, however, that (see (\ref{KnaufP})) $M_k$ is now always of the form $A_0 M_{k-1}$.  Thus, setting $x=0$ (the ``canonical" case), one sums over {\it all} Farey denominators, given here by $b+d$, at level $k$.  Letting $x=1$ to obtain the ``grand canonical" partition function thus corresponds to summing only over the ``new" denominators ($a+c+b+d$) at the next level. Now the Farey fractions at each level $k$ are composed of ``old" fractions that arose at lower levels and ``new" ones from level $k$.  Thus  the ``grand canonical" partition function at level $k$ can be written as a sum over ``canonical" partition functions at lower levels, i.~e.~over all ``canonical" chains of shorter length.  This is the opposite of the usual situation in statistical mechanics, and the reason why we put the names in quotes.

\section{Spin orientation transformations}\label{2}
In this section we consider the consequences of the spin-flip transformation generated by the matrix $P=\left({0\atop 1}{1\atop 0}\right)$. Specifically, we investigate the effects of $P$ on the partition function $Z_k(x,\beta)$ and some related functions useful in calculation expectation values. 

The action of $P$ on a function $f(x)$ is
\begin{equation}\label{Saction}f(x)|P=x^{-2\beta}f(1/x).
\end{equation}
Note that the matrix $P$ simply exchanges
the spin orientation, e.g.~the matrix $A_0$ and the matrix $A_1$ satisy
\begin{equation}\label{Aswitch}A_1=P\,A_0\,P.
\end{equation}
Since $P^2=1$, (\ref{Aswitch}) in fact implies that $A_0$ and $A_1$ are conjugate.
Note that a function $f(x)$ satisfying (\ref{Saction}) (i.e.~$f(x)=x^{-2\beta}f(1/x)$) can be called even,
since, using the substitution $e^y=x$ (recall that $x \ge 0$ herein) to define
$g(y)=e^{\beta}f(e^y)$, (\ref{Saction}) becomes $g(y)=g(-y)$.

Now our initial condition $Z_0(x,\beta)=(1+x)^{-2\beta}$ is easily seen to be even.
Consequently, for all $k\ge 1$, $x\in \mathbb R^+$
and $\beta \in \mathbb R^+_0$ the partition function 
$Z_{k}(x)$ is even
\begin{equation}\label{Zinvariance}Z_{k}(x)|P=(1+x)^{-2\beta}|(A_0+A_1)^kP=(1+x)^{-2\beta}
|P^2(A_0+A_1)^kP=Z_{k}(x).\end{equation}
In the last equality we used the evenness of our initial condition and the fact that the set 
of all terms in $(A_0+A_1)^k$ is the same
as the set $P(A_0+A_1)^kP$.  Note also that (\ref{fromgKnauf}) implies that the partition function $\tilde{Z}_k(x,\beta)$ is ``almost" even for $\beta<\beta_c$ and $k \to \infty$, since both $\tilde{Z}_k(x,\beta)-1$  and $\tilde{Z}_k(x,\beta)$ diverge in this limit, and the former is even.  Finally, (\ref{Saction}) and (\ref{Zinvariance}) show that  $Z_k(1)$, the ``grand canonical" partition function, is actually {\it invariant} under the spin-flip transformation.  This corresponds to the absence of odd-spin interactions in this model, as will be discussed below.

Now consider the terms in (\ref{rec}). Using the evenness of our
partition function we can write
$Z_{k-1}(x)|A_0=Z_{k-1}(x)|P\,A_0$ and $Z_{k-1}(x)|A_1=Z_{k-1}(x)|P\,A_1$. Thus
\begin{equation}\label{1ekv}
(1+x)^{-2\beta}Z_{k-1}\left(\frac{x}{1+x},\beta \right)
=x^{-2\beta}Z_{k-1}\left(\frac{1+x}{x},\beta \right)
\end{equation}
and
\begin{equation}\label{2ekv}Z_{k-1}(x+1,\beta)=
(1+x)^{-2\beta}Z_{k-1}\left(\frac{1}{1+x},\beta \right)\end{equation}
for all $k\ge 1$, $x\in \mathbb R^+$ and $\beta \in \mathbb R^+_0$.
Combining (\ref{1ekv}), (\ref{2ekv}) and (\ref{rec}) (which, as mentioned, also holds for $Z_k(x,\beta)$) gives us four different possible recursion formulas.
For instance
\begin{equation}\label{mrec1}Z_{k}(x)=(x+1)^{-2\beta}\left 
[Z_{k-1}\left ( \frac{x}{x+1}\right)+Z_{k-1}\left (\frac{1}{x+1}\right )\right ]
\end{equation}
which we will use in section \ref{3}. 

In addition we can see that the matrix $P$ can be put in
front of any matrix $A_0$ or $A_1$ in the expression $(1+x)^{-2\beta}|(A_0+A_1)^k$ without
changing the partition function $Z_{k}(x)$
(for example $(1+x)^{-2\beta}|(A_0+A_1)^l(P\,A_0+A_1)(A_0+A_1)^r=
(1+x)^{-2\beta}|(A_0+A_1)^k$ for any $k,l,r\ge 0$ such that $l+r+1=k$).
On the other hand if we put the matrix $P$ {\it after} any matrix $A_0$ 
or $A_1$ we get a new function. Let
\begin{equation}\label{Ufunction}Z_{k}^{_l\uparrow_r}(x)=\frac{1}{2}(1+x)^{-2\beta}|
(A_0+A_1)^l(A_0+A_1P)(A_0+A_1)^r\end{equation}
and
\begin{equation}\label{Dfunction}Z_{k}^{_l\downarrow_r}(x)=\frac{1}{2}(1+x)^{-2\beta}|
(A_0+A_1)^l(A_0P+A_1)(A_0+A_1)^r,
\end{equation}
with $l+r+1=k$.
Using (\ref{Aswitch}) we then have
\begin{equation}\label{Ufshort}Z_{k}^{_l\uparrow_r}(x)=Z_{l}(x)|A_0(A_0+A_1)^r
\end{equation}
and
\begin{equation}\label{Dfshort}Z_{k}^{_l\downarrow_r}(x)=Z_{l}(x)|A_1(A_0+A_1)^r.
\end{equation}
The arrows $\uparrow$ and $\downarrow$ refer to the interpretation of $A_0$ and $A_1$,
as up and down spins, respectively. Thus (\ref{Ufshort}) and (\ref{Dfshort})  motivate the notation in (\ref{Ufunction}) and (\ref{Dfunction}). In addition note that
\begin{equation}\label{DfUfsum}
Z_{k}^{_l\uparrow_r}(x)+Z_{k}^{_l\downarrow_r}(x)=Z_{k}(x).
\end{equation}
We will make use of these functions to calculate expectation values in section \ref{4}. 

Note that, as is often done in statistical mechanics, if we fix the spin at one position (or spins at several positions) and sum over the rest, as in (\ref{Ufshort}) or (\ref{Dfshort}), and then divide by the partition function, the result is an expectation value, since the ratio is the sum of the probabilities of all configurations with this spin (or these spins) fixed in the way specified.  This follows because each term in the partition function is the unnormalized probability of the corresponding spin configuration.  In statistical mechanics, expectation values involving more than one spin are sometimes referred to as ``correlations" or ``correlation functions", especially when one focuses on their dependence on the distance(s) between the spins.

We conclude with an observation which follows immediately
from (\ref{Ufshort}) and (\ref{Dfshort}). The probability of a spin up at
position $l+1$ from the left is equal to the
probability of a spin down at the same position for a model with different $x$.
It is easy to see that
\begin{equation}\label{updownrel}
\frac{Z_{k}^{_l\uparrow_r}(x)}{Z_{k}(x)}=\frac{Z_{k}^{_l\downarrow_r}(x)|
P}{Z_{k}(x)}=\frac{x^{-2\beta}Z_{k}^{_l\downarrow_r}(1/x)}{Z_{k}(x)|P}=
\frac{Z_{k}^{_l\downarrow_r}(1/x)}{Z_{k}(1/x)}
\end{equation} 
for all $k\ge 1$, $x\in \mathbb R^+$ and $\beta \in \mathbb R^+_0$. Note that
for $x=1$ these probabilities are equal. 
For other values of $x$, since the magnetization (which is essentially the probability of spin up minus the probability of spin down) is zero, the up and down spins probabilities become equal when 
$l$ and $r$ are sent to infinity (see section \ref{4}).

\section{Connection to the transfer operator}\label{3}
In this section we demonstrate a direct and simple connection between the partition function $Z_k(x,\beta)$ and a transfer operator for a model of intermittency in dynamical systems associated with the Farey fractions. This connection had already been noticed in \cite{K-o1}, but in a less direct setting. Our new result allows us to prove that the free energy (cf.~(\ref{freeE})), which is given by the largest eigenvalue of the operator, is independent of $x$, and draw other conclusions as well.  In particular, the spectrum of this operator has been determined by Prellberg \cite{P-s}, and it follows from his results that there is a second-order phase transition for all $x$. 

To begin, consider the Farey tree, which is generated by the Farey map acting on the unit interval $[0,1]$, or more precisely, on the point $x=1/2$.  It consists, at each level, of a subset of the Farey fractions.  (For more details on these matters, see \cite{F} and \cite{FK}).  The Farey map is defined as
\begin{equation}\label{fm}
f(x)=\left\{\begin{array}{c}
f_0(x)=x/(1-x)\;,\quad\mbox{if $0\leq x\leq 1/2$,}\\
f_1(x)=(1-x)/x\;,\quad\mbox{if $1/2<x\leq 1$\;.}
\end{array}\right.
\end{equation}
We denote the inverses by $F_0(x)={f_0}^{-1}(x)=x/(1+x)$ and $F_1(x)={f_1}^{-1}(x)=1/(1+x)$.
The associated Ruelle-Perron-Frobenius transfer operator is then formally given by (note the resemblance to (\ref{mrec1}))
\begin{eqnarray}\label{to}
{\cal K}_\beta\;\phi(x)&=&
|{F_0}'(x)|^\beta\phi(F_0(x))+|{F_1}'(x)|^\beta\phi(F_1(x))\nonumber\\
&=&\frac1{(1+x)^{2\beta}}\left[\phi\left(\frac x{1+x}\right)+\phi\left(\frac1{1+x}\right)\right]\;.
\end{eqnarray}
Therefore, the $k$-fold iterated operator ${\cal K}_{\beta}^k\;\varphi(x)$ consists of $2^k$ terms of the form
\begin{equation}\label{term}
|(F_{\tau_1}\circ F_{\tau_2}\circ\ldots\circ F_{\tau_k})'(x)|^\beta
\varphi(F_{\tau_1}\circ F_{\tau_2}\circ\ldots\circ F_{\tau_k}(x))
\end{equation}
with $\tau_j\in\{0,1\}$. 
As we are dealing with iterations of M\"obius transformations of the
form $\frac{ax+b}{cx+d}$ with determinant $\pm1$, we can alternatively 
consider multiplication of the associated matrices. We find for instance
\begin{eqnarray}\label{sc}
{\cal K}_{\beta}^k1(x)=\sum_{\{\tau_j\}}(cx+d)_{\{\tau_j\}}^{-2\beta}=
\sum_{i=1}^{2^k}1(x)|\tilde{M_i},
\end{eqnarray}
where $c$ and $d$ are just the bottom left and right entries, respectively, of the matrix
\begin{equation}\label{nm}
\tilde{M_i}=\prod_{j=1}^{k}F_{\tau_j(i)}\quad\text{where}\quad
F_0=\begin{pmatrix}1&0\\1&1\end{pmatrix}
\quad\text{and}\quad
F_1=\begin{pmatrix}0&1\\1&1\end{pmatrix}\;.
\end{equation}
Note that $A_0=F_0$ and $F_1=PA_1$.

When we apply ${\cal K}_{\beta}$ to the constant function $1(x)$ we
obtain $2(1+x)^{-2\beta}$. That is exactly twice the initial condition of the partition
function  
$Z_{k}(x)$ (see (\ref{defofpartfunction})). In addition ${\cal K}_{\beta}$ increases the level $k$
of the partition function $Z_{k}(x)$ by one as follows from (\ref{mrec1})
and (\ref{to}).
Thus
\begin{equation}\label{a1}
{\cal K}_{\beta}^k1(x)=2Z_{k-1}(x).
\end{equation}
(\ref{a1}) is one of our main results.  A connection of this type follows from \citep{C-K} and \cite{K-o1}, but it is less direct, and not valid for all $x$ values.  Next, we consider some of the consequences of (\ref{a1}).

First, we note that for $x=0$, (\ref{a1}) connects the Knauf model (\ref{KnaufP}) 
and the transfer operator ${\cal K}_{\beta}$:
\begin{equation}\label{Kpart}
{\cal K}_{\beta}^k1(x)|_{x=0}=2\,Z_{k-1}^K(2\beta).
\end{equation}
Now  \citep{K-o1}
 defines an operator $\tilde{{\cal C}}(2\beta)$  whose non-degenerate leading eigenvalue $\lambda(\beta)$ gives the free energy of the  ``grand canonical" partition function $Z_k(x=1,\beta)$ and the ``canonical" case $Z_k(x=0,\beta)$
as in (\ref{freeEF}) below. It also connects the largest eigenvalue of $\tilde{{\cal C}}(2\beta)$ with the largest eigenvalue of the equation
\begin{equation}\label{LewisThreeterm}
\lambda (\beta) f(x)=f(x+1)+x^{-2\beta}f(1+1/x),
\end{equation}
which is directly related to (\ref{to}). In fact the proof uses a Taylor series expansion of $\phi(x)$ (in (\ref{to})) at $x=1$. However,  the connection of the partition functions and (\ref{to}) or (\ref{LewisThreeterm}) in  \citep{K-o1} is not so direct. 

A connection between the spectrum of the operator $\tilde{{\cal C}}(2\beta)$ and the spectrum of (\ref{to}) can be made, but the situation is complicated (in part because the spectrum determined in \cite{P-s} is on the space of functions of bounded variation), and not really germane to our purpose here, and so will be omitted.

Next consider (\ref{Kpart}) for $\beta>\beta_c=1$. In 
that case, one has
\begin{equation}\label{limKpart}
\lim_{k\to\infty}{\cal K}_{\beta}^k1(x)|_{x=0}=\lim_{k\to\infty}2\,Z_{k-1}^K(2\beta)
=2\frac{\zeta (2\beta -1)}{\zeta (2\beta)},
\end{equation}
where $\zeta$ is the Riemann zeta-function (the second equality is shown in \citep{K}).  This result has not appeared previously, to our knowledge.

For $\beta<\beta_c=1$, the leading eigenvalue $\lambda (\beta)>1$ of ${\cal K}_{\beta}$ \cite{P-s} is non-degenerate and belongs to the discrete spectrum.
Since the corresponding eigenvector is of definite sign, it has a non-zero projection onto $1(x)$.  Thus we can define $a(x,\beta)$ as 
\begin{equation}\label{lim}
a(x,\beta)=\lim_{k\to\infty}\frac{Z_k(x,\beta)}{\lambda^k(\beta)}<\infty.
\end{equation}
Note that since the spectrum of ${\cal K}_{\beta}$ is independent of $x$, the free energy 
\begin{equation}\label{freeE}
f(\beta) :=\frac{-1}{\beta} \lim_{k\rightarrow\infty}\frac{\ln
Z_k(x,\beta)}{k} = \frac{-1}{\beta} \ln \lambda(\beta)
\end{equation}
depends only on the inverse temperature $\beta$  \citep{K-o1,FK}.
Thus (as we have already noted for $x = 0$ in  {\citep{FK}) the phase transition is second-order for all $x\ge 0$. 
This follows from the result of Prellberg \citep{P-S,Diss},
\begin{equation}\label{prellas}
\lambda (\beta) =c\frac{\beta-1}{\ln (1-\beta)}[1+o(1)], \quad \beta \to 1^-,
\end{equation}
where $c>0$ (for more discussion about the phase transition see  \citep{FK}).

Now since all terms in $Z_k(x,\beta)$ are positive, and the matrix $A_0^{k+1}$ is included (see (\ref{defofpartfunction})), one has (recall that $x \ge 0$) 
\begin{equation}\label{inequalPar}
(1+(k+1)x)^{-2\beta}\le Z_k(x,\beta)\le Z_k(0,\beta)=Z_k^K(2\beta).
\end{equation}
Thus, since the Knauf free energy vanishes for $\beta\ge\beta_c$, so must the free energy obtained from $Z_k(x,\beta)$.
Furthermore, since the leading eigenvalue of ${\cal K}_{\beta}$ is $\lambda(\beta)=1$ for all $\beta\ge\beta_c$ we can write for all temperatures
\begin{equation}\label{freeEF}
f(x,\beta) =\frac{-1}{\beta}\ln \lambda(\beta).
\end{equation}

We have shown elsewhere \cite{FK} that the free energy of the
Knauf model, the Farey tree model and the Farey fraction spin chain of Kleban and \"{O}zl\"{u}k  are the same for all temperatures  and are also given by (\ref{freeEF}).

Note that the leading eigenvalue changes its character at the critical point. 
Above the critical temperature it belongs to a discrete 
spectrum and below the critical temperature
it is the upper limit of the continuous spectrum (for more details about the spectrum see \cite{P-s}). 

The sub-leading eigenvalue in the spectrum  is equal to one for $\beta \le \beta_c$. 
This is consistent with our previous results in \citep{FK1} based on scaling and renormalization
group arguments. For a one-dimensional system the scaling arguments provide the relation between the singular part of free energy $f_s$ and correlation length $\xi$
\begin{equation}\label{sacl}
f_s\propto \frac{1}{\xi}.
\end{equation}
(The correlation length is essentially the distance over which the spin-spin correlation function varies, i.~e.~is not constant.)  If we assume that our partition function goes as 
\begin{equation}\label{asfe}
Z_k(x,\beta)=\lambda^k a(x)+\lambda_1^k a_1(x)+\ldots ,
\end{equation}
we obtain, using (\ref{freeE}),
\begin{equation}\label{freesingresult}
f_s\propto \ln\lambda,
\end{equation}
and from the definition of the correlation length
\begin{equation}\label{correrelat}
\xi=\frac{C}{\ln( \lambda /\lambda_1)}
\end{equation}
where C is a positive constant. This implies that the sub-leading eigenvalue $\lambda_1(\beta)=1$ for $\beta\le\beta_c$,
consistent with Prellberg's results.

In addition, note that from (\ref{lim}) and the evenness of $Z_k$, it follows that the eigenfunction $a(x,\beta)$ is even 
\begin{equation}\label{ae}
a(x,\beta)=x^{-2\beta}a(1/x,\beta).
\end{equation}
Using this fact and (\ref{to}) we can write
\begin{equation}\label{eign}
\lambda(\beta)a(x,\beta)=a(x+1,\beta)+(1+x)^{-2\beta}a\left (\frac{x}{x+1},\beta\right ).
\end{equation}
Note that, as remarked in \cite{K-o1} (and using the evenness of $a(x,\beta)$), (\ref{eign}) is a generalization of the Lewis three-term equation, which has been extensively studied in number theory in the context of the Selberg $\zeta$-function (\cite{LZ, L-Z}).  In the Lewis case, solutions with $\lambda=1$ are of interest, and $\beta$ may be complex.

Applying (\ref{ae}) and (\ref{eign}) with  $x=0$ and $x=1$ we obtain
\begin{equation}\label{a2}
a(1,\beta)=(\lambda(\beta)-1)a(0,\beta),
\end{equation}
and
\begin{equation}\label{a3}a(2,\beta)=\frac{\lambda(\beta)}{2}a(1,\beta)=\frac{\lambda(\beta)}{2}(\lambda(\beta)-1)
a(0,\beta),\end{equation}
respectively. We will make extensive use of (\ref{a2}) and (\ref{a3}) below.

\section{Expectation values-preliminaries}\label{4}
In this section we consider various spin expectation values  for Knauf spin chains. (The remarks just below (\ref{DfUfsum}) define these quantities.) Making use of the spin flip behavior and recursion relations proved above, we obtain a few results, but our main purpose is to set the stage for the expectation value calculations of the next section. 

First, consider the expectation value for spin up 
\begin{equation}\label{expvalup}
\langle \underbrace {\ldots}_{l} \uparrow
\underbrace {\ldots}_{r} \rangle_x :=\frac{Z_{k}^{_l\uparrow_r}(x)}{Z_{k}(x)},
\end{equation}
and similarly for spin down.
By using (\ref{Zinvariance}), (\ref{Ufshort}) and (\ref{DfUfsum}) we find 
\begin{eqnarray}\label{onespin}\langle \underbrace {\ldots}_{l} \uparrow
\underbrace {\ldots}_{r} \rangle_x 
&=& 
\frac{Z_{l}(x)|A_0(A_0+A_1)^r}
{Z_{l}(x)|A_0(A_0+A_1)^r+Z_{l}(x)|A_0(A_0+A_1)^rP}.
\end{eqnarray}
We now relate the two terms in the denominator, at least for some values of $x$.
We already know from (\ref{updownrel}) that for $x=1$ these terms are equal. There is a simple explanation for 
this. Multiplying any  matrix  $M_i$ by $P$ on the right just exchanges its columns, and (\ref{def2}) is clearly invariant under exchange of columns for $x=1$.
Thus the probability to find spin up (or down) at any location on the spin chain with $x=1$
is 
\begin{equation}\label{onespinxone}
\langle \underbrace {\ldots}_{l} \uparrow
\underbrace {\ldots}_{r} \rangle_{x=1} =\langle \underbrace {\ldots}_{l} \downarrow
\underbrace {\ldots}_{r} \rangle_{x=1} =\frac{1}{2}\; \cdot
\end{equation}
Thus, in this case, due to the spin-flip symmetry, there are no finite size or edge effects at all (the result is valid for all $l,r \geq 0$).  The situation is very different for $x=0$, as we will see.

Although our spin chains are defined in terms of matrices, one can also investigate their Hamiltonians.  Generally, these are not very useful, since they include long-range many-body interactions between the spins (see (\cite{K}) or (\cite{K-O}) for definitions and explanations of these matters). However, it is known that  for the ``grand canonical" spin chain, all interactions are even and ferromagnetic (i.~e.~favoring aligned spins) \cite{K}. Therefore any expectation value involving an odd number of spins must vanish. Since $x=1$ corresponds to this spin chain,  (\ref{onespinxone}) is exactly what one expects.

Now consider $x=0$ (the Knauf model of (\ref{KnaufP})). The partition function at level $k$ is
\begin{equation}\label{xzero}
Z_k(0)=(Z_{l}(x)|A_0(A_0+A_1)^r+Z_{l}(x)|A_0(A_0+A_1)^rP)|_{x=0}=2Z_{k}^{_l\uparrow_r}(0)+Z_l(1)-Z_l(0).
\end{equation}
This result may be proven directly from the structure of the Farey fractions together with the action of the matrix $P$. 
However, we will show it by using (\ref{Ufshort}) and (\ref{Dfshort}).  
First, note that $Z_{k-r}^{_l\uparrow_0}(0)=
Z_{l}(0)$ and $Z_{k-r}^{_l\downarrow_0}(0)=
Z_{l}(1)$.  Next,  express (\ref{Ufshort}) as \begin{equation}\label{recUP}
Z_{k}^{_l\uparrow_r}(x)=
(1+x)^{-2\beta}Z_{k-1}^{_l\uparrow_{r-1}}\left(\frac{x}{1+x} \right)
+Z_{k-1}^{_l\uparrow_{r-1}}(x+1).
\end{equation}  
Now for $x=0$ (\ref{recUP}) becomes 
\begin{equation}\label{recUPx0}
Z_{k}^{_l\uparrow_r}(0)=
Z_{l}(0)
+\sum_{i=l+1}^{k-1}Z_{i}^{_l\uparrow_{i-l-1}}(1).
\end{equation}  
Similarly we find
\begin{equation}\label{recDPx0}
Z_{k}^{_l\downarrow_r}(0)=
Z_{l}(1)
+\sum_{i=l+1}^{k-1}Z_{i}^{_l\downarrow_{i-l-1}}(1).
\end{equation}  
Adding the above expressions (see(\ref{DfUfsum})) and using the fact that $Z_{k}^{_l\uparrow_r}(1)=Z_{k}^{_l\downarrow_r}(1)$ (see (\ref{updownrel})) leads to (\ref{xzero}).

By making use of (\ref{onespin}) and (\ref{xzero}), the expectation value for spin up at $x=0$ can be written as \begin{eqnarray}\label{1spinzero} 
\langle \underbrace {\ldots}_{l} \uparrow
\underbrace {\ldots}_{r} \rangle_{x=0} 
&=& \frac{1}{2-K},
\end{eqnarray}
where 
\begin{equation}\label{defA}
K=\frac{Z_{l}(0)-Z_{l}(1)}{Z_{l}(x)|A_0(A_0+A_1)^r|_{x=0}}.
\end{equation}
 Now $Z_{l}(x)|A_0(A_0+A_1)^r|_{x=0}\ge Z_{l}(0) >  Z_{l}(1) > 0$ for all $l,r\ge 0$ and $\beta > 0$.
The first inequality follows immediately from the fact that the sum $Z_{l}(x)|A_0(A_0+A_1)^r|_{x=0}$ of positive terms includes the term $Z_{l}(x)|A_0^{r+1}|_{x=0}=Z_{l}(0)$.
The second inequality follows directly from the monotonicity (in $x$) of  $Z_{l}(x)$.  Therefore  $0 \le K \le 1$, where $K=0$ can occur if the denominator of (\ref{defA}) diverges.  (This happens when $r \to \infty$, see (\ref{expvalueleft}) below.) Thus the spin at any position for temperature $T < \infty$  has, in general, greater probability to be up than down 
\begin{eqnarray}\label{1spinzeroCON} 
\langle \underbrace {\ldots}_{l} \uparrow
\underbrace {\ldots}_{r} \rangle_{x=0} 
& \ge & \langle \underbrace {\ldots}_{l} \downarrow
\underbrace {\ldots}_{r} \rangle_{x=0},
\end{eqnarray}
where it should be realized that equality only holds in the special case $K=0$.

In the ``normal" situation, i.~e.~when equality does not hold, (\ref{1spinzeroCON}) may be regarded as an effect of the ``hidden" spin up on the left, i.e.~the initial condition $(1+x)^{-2\beta}=1(x)|A_0$, which breaks spin-flip symmetry.  A slightly different point of view involves the spin interactions.  For $x=0$, i.~e.~the (canonical) Knauf model, these are all ferromagnetic and include  terms  with an odd number of spins \citep{K, C-Kn}.   (\ref{1spinzeroCON}) shows that the odd interactons can be sufficient to favor an up spin. The  interactions also give rise to some  rather subtle effects in certain other expectation values, as we will see below.

Now we consider the two-spin correlation function. Let
\begin{eqnarray}\label{twospins}
\langle \underbrace {\ldots}_{l} \uparrow \underbrace {\ldots}_{n}\uparrow
\underbrace {\ldots}_{r} \rangle_x 
&=& 
\frac{Z_{k}^{_l\uparrow_n}(x)|A_0(A_0+A_1)^r}
{Z_{k+r+1}(x)},
\end{eqnarray}
where as before $k=l+n+1$.

The partition function $Z_{k+r+1}(x)$ for a spin chain of length $l+n+r+2$ can be divided into four terms (corresponding to the four possible configurations of two spins)
$$Z_{k}^{_l\uparrow_n}(x)|A_i(A_0+A_1)^r$$ and $$Z_{k}^{_l\downarrow_n}(x)|A_i(A_0+A_1)^r$$ where $i\in \{0,1\}$. 
Using the matrix $P$ (see (\ref{updownrel})) gives
\begin{equation}\label{relations}
Z_{k}^{_l\uparrow_n}(x)|A_i(A_0+A_1)^r=Z_{k}^{_l\downarrow_n}(x)|A_{i+1 ({\rm mod}\, 2)}(A_0+A_1)^rP.
\end{equation}
Now  (\ref{onespinxone})  shows that for $x=1$ each spin has equal probability to be up or down without
any edge or finite size effects (i.e.~for any $l,$ $r\in \mathbb Z_0^+$). Thus we can expect that e.g.~the expectation value for two spins up is the same as for two spins down.
In fact (\ref{twospins}) and (\ref{relations}) give immediately
\begin{eqnarray}\label{twospinsxoneuu}
\langle \underbrace {\ldots}_{l} \uparrow \underbrace {\ldots}_{n}\uparrow
\underbrace {\ldots}_{r} \rangle_{x=1}=\langle \underbrace {\ldots}_{l} \downarrow \underbrace {\ldots}_{n}\downarrow
\underbrace {\ldots}_{r} \rangle_{x=1} 
\end{eqnarray}
and
\begin{eqnarray}\label{twospinsxoneud}
\langle \underbrace {\ldots}_{l} \uparrow \underbrace {\ldots}_{n}\downarrow
\underbrace {\ldots}_{r} \rangle_{x=1}=\langle \underbrace {\ldots}_{l} \downarrow \underbrace {\ldots}_{n}\uparrow
\underbrace {\ldots}_{r} \rangle_{x=1}
\end{eqnarray}
where  $l,$ $n,$ $r\in \mathbb Z_0^+$.

In the case of one spin (\ref{onespinxone}) shows that the expectation value does not change under translation of the spin. The two spin expectation 
value is not translationally invariant but it does have the following symmetry 
\begin{eqnarray}\label{twospinsflip}
\langle \underbrace {\ldots}_{l} \uparrow \underbrace {\ldots}_{n}\downarrow
\underbrace {\ldots}_{r} \rangle_{x=1}=\langle \underbrace {\ldots}_{r} \downarrow \underbrace {\ldots}_{n}\uparrow
\underbrace {\ldots}_{l} \rangle_{x=1}.
\end{eqnarray}
This follows on rewriting the l.~h.~s.~of (\ref{twospinsflip}) as
\begin{eqnarray} \nonumber
Z_{k}^{_l\uparrow_n}(x)|A_1(A_0+A_1)^r
&=&
(1+x)^{-2\beta}|(A_0+A_1)^lA_0(A_0+A_1)^nA_1(A_0+A_1)^r \\  \label{numerator}
&=&
\sum_{i=1}^{2^{l+n+r}}[(a+c)x + b+d]_i^{-2\beta}, 
\end{eqnarray}
where $a$, $b$, $c$, $d$ are entries of the $i$th matrix from the set $(A_0+A_1)^lA_0(A_0+A_1)^nA_1(A_0+A_1)^r$. Thus for $x=1$ the sum does 
not change under matrix transposition and we get (\ref{twospinsflip}).  This result also follows from the proof that the interactions in the ``grand canonical" spin chain also have the symmetry (\ref{twospinsflip}) (Lemma 4.8 in \cite{K}).

\section{Expectation values-results}\label{5}
In this section we calculate several spin expectation values  for Knauf spin chains. These results are all new, and are the first calculations of such quantities, to our knowledge.  We find that they are expressed as simple functions of the free energy $f$ (or correlation length $\xi$).

The methods used in the previous section (and this one as well!) are only of use when one can come up with a finite, closed set of equations.  For the expectation value of a general set of spins, this is not the case.  Since the matrices representing the spins operate ``from the right" most of our results are expectation values involving a finite number of spins fixed at or at a finite distance from the right hand side of the spin chain. 

First note that by (\ref{onespinxone}), at $x=1$, the spin expectation value has no edge or finite size effects. 
Thus, allowing $l\to \infty$ and $r\to \infty$, a spin up (down) still has probability one half. By contrast, (\ref{1spinzero}) shows that there may be such effects for
 $x=0$.  This indeed occurs, as we now proceed to demonstrate. 

In order to see the edge effect at the right side of an infinitely long chain  we go back to (\ref{xzero}) and let $l\to \infty$ . 
Using (\ref{lim}) and the properties of the eigenfunction $a(x)$ we get
\begin{equation}\label{limxzero}
\lambda^{r+1}a(0)=2\,a(x)|A_0(A_0+A_1)^r|_{x=0}+a(1)-a(0)
\end{equation}
for all $r\ge 0$. Thus, we can write the expectation value for a spin $r+1$ from the right of the infinitely long chain using (\ref{a2})
\begin{equation}\label{expvalueright}
\langle \underbrace {\ldots}_{\infty} \uparrow
\underbrace {\ldots}_{r} \rangle_{x=0}=\frac{1}{2}(1+\frac{2-\lambda}{\lambda^{r+1}}),
\end{equation}
where the eigenvalue $\lambda(\beta) \in (1,2]$ for $\beta \in [0,\beta_c)$.  Note that a similar expression for the spin down expectation value follows since their sum must be one.  

Now recall (see (\ref{freeEF}))} that $\lambda$ is given directly in terms of the free energy via $\lambda = e^{-\beta f}$.  The free energy is a non-increasing function of the temperature for $\beta \le \beta_c$.  Hence, for any fixed $r$, the expectation value (\ref{expvalueright}) decreases monotonically to $1/2$ as $T \to \infty$.  In fact, all our results are consistent with a product distribution in this limit, i.~e.~the probability of a given spin being up or down is $1/2$.  Note also that $\lambda$ may also be expressed in terms of the correlation length $\xi$ (see (\ref{correrelat}) and recall that $\lambda_1 = 1$ for $\beta \le \beta_c$).

From a physical point of view, it is also interesting to compare  (\ref{expvalueright}) and   (\ref{1spinzeroCON}), which, as mentioned, may be attributed to long-range  interactions between an odd number of spins. (\ref{expvalueright}) shows that their effects are felt even infinitely far from the initial (``hidden") up spin.  This is particularly interesting, since \cite{K} proves that even though there are odd (ferromagnetic) interactions at $x=0$, any individual interaction term vanishes in the limit of an infinitely long chain.  Thus (\ref{expvalueright}) shows that certain cumulative effects of the odd interactions remain in this limit, even though each individual interaction goes to zero.

It is also of note that (\ref{expvalueright}), as well as various expressions that we will derive shortly, give expectation values as simple polynomials in $\lambda$, which is itself exponential in the free energy $f$ or correlation length $\xi$, as mentioned.

We can use (\ref{expvalueright}) at the critical temperature (where $\lambda (\beta_c)=1$) by taking the limit $\beta\to\beta_c$. 
Then the probability 
of a spin up is 1 for any finite distance $r$ from the right (this can also be shown directly from (\ref{1spinzero}) since $K\to 1$ when $\beta\to\beta_c$ and then $l\to \infty$).
On the other hand for any $\beta<\beta_c$ the spin up (or down!) probability goes to one half as $r\to \infty$.

Note that (\ref{expvalueright}) also gives the right edge correlation length $\xi_r$ as
\begin{equation}\label{edgecorrelation}
\xi_r=\frac{1}{\ln \lambda}=\frac{1}{f_s}.
\end{equation}
This equation directly relates edge and bulk behavior. Since the bulk correlation length $\xi\propto \frac{1}{f_s}$ (see (\citep{FK1})),
\begin{equation}\label{edgebulkcorrelation}
\xi_r\propto\xi\propto \frac{\ln \epsilon}{\epsilon}
\end{equation}
as $\beta\to\beta_c$, where $\epsilon=\frac{\beta_c}{\beta}-1$.

Now consider the limit $r\to \infty$, keeping $l$ finite.   Using (\ref{xzero}) we can write
\begin{equation}\label{rightlimitval}
\lim_{r\to\infty}\frac{Z_{l}(x)|A_0(A_0+A_1)^r|_{x=0}}{\lambda^r}=\frac{\lambda^{l+1}}{2}\,a(0)
\end{equation}
for any $\lambda >1$ (i.e.~$\beta<\beta_c$). From ((\ref{inftwospinsinf}) below we see that $0<a(0)<\infty$ for $\lambda \in (1,2]$. 
Using (\ref{1spinzero}) and (\ref{rightlimitval}) we then obtain
\begin{equation}\label{expvalueleft}
\langle \underbrace {\ldots}_{l} \uparrow
\underbrace {\ldots}_{\infty} \rangle_{x=0}=\frac{1}{2}
\end{equation}
for all $l\ge 0$ and $\beta<\beta_c$. Thus the left edge effects on one spin vanish. From a physical point of view, this is quite interesting.  The ``hidden" spin up on the left, or equivalently the long-range odd ferromagnetic interactions \citep{K} have no effect in an infinite chain when the spin in question is only a finite distance from the ``hidden" spin.  By contrast, when it is infinitely far away but at a finite distance from the right edge, there is an effect (see   (\ref{expvalueright})).  However, we will see that this effect is removed if one fixes spins on the right hand end of the chain in certain specific configurations.

Next we consider the two spin correlation function, in the limit where the left part of the spin chain goes to infinity.  Using (\ref{Ufshort}), (\ref{lim}) and (\ref{twospins})
we find
\begin{eqnarray}\label{inftwospins}
\langle \underbrace {\ldots}_{\infty} \uparrow \underbrace {\ldots}_{n}\uparrow
\underbrace {\ldots}_{r} \rangle_x 
&=& 
\frac{a(x)|A_0(A_0+A_1)^nA_0(A_0+A_1)^r}
{\lambda^{n+r+2}a(x)}.
\end{eqnarray}
It is convenient to define two functions of $x$ and $\beta$ (where, as for $a(x)$, we do not explicitly indicate the $\beta$ dependence),
\begin{equation}\label{Updef}  
U_n(x)=a(x)|A_0(A_0+A_1)^{n}
\end{equation}
for spin up and similarly for spin down
\begin{equation}\label{Dndef}
D_n(x)=a(x)|A_1(A_0+A_1)^{n}.
\end{equation}
Clearly for all $n\ge 0$ and $0\le\beta<\beta_c=1$ 
\begin{equation}\label{UDrelation1}
U_n(x)+D_n(x)=\lambda^{n+1}a(x).
\end{equation}
Using (\ref{Aswitch}) and (\ref{ae}) 
\begin{equation}\label{UDrelation2}
U_n(x)=x^{-2\beta}D_n(1/x).
\end{equation}

Now return for a moment to the one-spin expectation value.  We can write 
\begin{equation}\label{Uap}
\langle \underbrace {\ldots}_{\infty} \uparrow
\underbrace {\ldots}_{n} \rangle_x =\frac{U_n(x)}{U_n(x)+D_n(x)}.
\end{equation}
For $x=1$ it immediately follows from (\ref{UDrelation2}) that
\begin{equation}\label{Uapx0}
\langle \underbrace {\ldots}_{\infty} \uparrow
\underbrace {\ldots}_{n} \rangle_{x=1} =\frac{1}{2},
\end{equation}
as already shown by (\ref{onespinxone}).
Note also that (\ref{expvalueright}), for one spin at $x=0$, follows from equation (\ref{limxzero})
which we rewrite as
\begin{equation}\label{Ua1}
U_n(0)=\left(\frac{1}{2}(\lambda^{n+1}-\lambda)+1\right )a(0),
\end{equation} 
and similarly 
\begin{equation}\label{Da1}
D_n(0)=\left(\frac{1}{2}(\lambda^{n+1}+\lambda)-1\right )a(0).
\end{equation}

Now return to equation (\ref{inftwospins}) for $r \to \infty$. First we rewrite it as
\begin{eqnarray}\label{inftwospinsmod}\nonumber
\langle \underbrace {\ldots}_{\infty} \uparrow \underbrace {\ldots}_{n}\uparrow
\underbrace {\ldots}_{r} \rangle_x 
&=& 
\frac{U_n(x)|A_0(A_0+A_1)^r}{\lambda^{n+r+2}a(x)} \\ 
&=&
\frac{\sum U_n(\frac{ax+b}{cx+d}) (cx+d)^{-2\beta}}{\lambda^{n+r+2}a(x)},
\end{eqnarray}
where the sum has $2^r$ terms, and $a$, $b$, $c$ and $d$ are from $A_0M_r=\left({a\atop c}{b\atop d}\right)$.  Note that we start with the matrix $A_0$ and thus $\frac{ax+b}{cx+d}\le 1$ for all $x\in \mathbb R^+_0$. 
Now
\begin{eqnarray}\label{maxminU}
U_n(x)=\lim_{k\to \infty}\sum_{i=1}^{2^{k+n}} \frac{(cx+d)^{-2\beta}_i}{\lambda(\beta)^k},
\end{eqnarray}
where $\left({a_i\atop c_i}{b_i\atop d_i}\right)\in \{ A_0(A_0+A_1)^kA_0(A_0+A_1)^n \}$. Since $c_i, d_i >0$ for all $i$, it follows that $U_n(x)$ is non-increasing with $x$.
Thus we can write
\begin{eqnarray}\label{inftwospinsnove}
\frac{U_n(1)Z_r(x)}{\lambda^{n+r+2}a(x)}\le
\langle \underbrace {\ldots}_{\infty} \uparrow \underbrace {\ldots}_{n}\uparrow
\underbrace {\ldots}_{r} \rangle_x \le
\frac{U_n(0)Z_r(x)}{\lambda^{n+r+2}a(x)}
\end{eqnarray}
for all $r\ge 0$ and all $x\in \mathbb R^+_0$. In the limit $r \to \infty$ we get, using (\ref{Ua1}) and 
$U_n(1)=\lambda^{n+1}a(1)/2$ (see (\ref{UDrelation1}) and (\ref{UDrelation2}))
\begin{eqnarray}\label{inftwospinsinf}
(\lambda -1)\frac{a(0,\lambda)}{2\lambda}\le
\langle \underbrace {\ldots}_{\infty} \uparrow \underbrace {\ldots}_{n}\uparrow
\underbrace {\ldots}_{\infty} \rangle_x \le
\left ( 1+\frac{2-\lambda}{\lambda^{n+1}}  \right) \frac{a(0,\lambda)}{2\lambda}.
\end{eqnarray} 
Physically, since the correlation length $\xi=\frac{1}{\ln \lambda}$, the $n$-dependence of the upper bound in (\ref{inftwospinsinf})
is what one expects for the correlation function itself. We have not been able to prove this, however.

Now we calculate some results at the right hand edge, i.e.~for finite $r$, with $x=0$ (the ``canonical" case).
When $r=0$ we have from (\ref{Ua1}) and  (\ref{inftwospinsmod})
\begin{eqnarray}\label{UUinftwospinssmallr0}
\langle \underbrace {\ldots}_{\infty} \uparrow \underbrace {\ldots}_{n}\uparrow \rangle_{x=0} 
&=& 
\frac{U_n(0)}{\lambda^{n+2}a(0)}=\left ( 1+\frac{2-\lambda}{\lambda^{n+1}}  \right) \frac{1}{2\lambda}.
\end{eqnarray}
Similarly 
\begin{eqnarray}\label{DUinftwospinssmallr0}
\langle \underbrace {\ldots}_{\infty} \downarrow \underbrace {\ldots}_{n}\uparrow \rangle_{x=0} 
&=& 
\frac{D_n(0)}{\lambda^{n+2}a(0)}=\left ( 1-\frac{2-\lambda}{\lambda^{n+1}}  \right) \frac{1}{2\lambda},
\end{eqnarray}
as well as (see (\ref{a2}))
\begin{eqnarray}\label{UDinftwospinssmallr0}
\langle \underbrace {\ldots}_{\infty} \uparrow \underbrace {\ldots}_{n}\downarrow \rangle_{x=0} 
&=& 
\frac{U_n(1)}{\lambda^{n+2}a(0)}=\frac{\lambda-1}{2\lambda},
\end{eqnarray}
and
\begin{eqnarray}\label{DDinftwospinssmallr0}
\langle \underbrace {\ldots}_{\infty} \downarrow \underbrace {\ldots}_{n}\downarrow \rangle_{x=0} 
&=& 
\frac{D_n(1)}{\lambda^{n+2}a(0)}=\frac{\lambda-1}{2\lambda}.
\end{eqnarray}
It is easy to see, for instance, that the sum of (\ref{UUinftwospinssmallr0}) and (\ref{UDinftwospinssmallr0}) is the same as (\ref{expvalueright}), and that  (\ref{UUinftwospinssmallr0}) and (\ref{DUinftwospinssmallr0}) sum to (\ref{expvalueright}) with $r=0$. Similar checks verify other sums of the four equations just above.  Further, since the interactions are ferromagnetic, (\ref{UUinftwospinssmallr0}) must be the largest of the three, and this is easily verified as well (recall that $1 \le \lambda \le 2$).

It is interesting  that both (\ref{UDinftwospinssmallr0}) and (\ref{DDinftwospinssmallr0}) are completely independent of the spin separation $n$, and equal to each other for $\beta < \beta_c$.  Thus a down-spin at the right hand edge completely cancels the lack of spin symmetry seen in (\ref{expvalueright}), and does so for all $\beta < \beta_c$.  (In fact this holds for $\beta \ge \beta_c$ as well, since all spins are up in an infinite chain, so both  (\ref{UDinftwospinssmallr0}) and (\ref{DDinftwospinssmallr0}) vanish.)  We comment further on this after deriving some more general results. 

The results in the paragraph above are based on our knowledge of $U_n(x)$ and $D_n(x)$ at the two values $x=0$ and $x=1$.
It is easy to find generalizations. We need combinations of spins for which the corresponding product of matrices $A_0$
and $A_1$ has $b=0$, so that $x=0$ is preserved,  or $b=1$ and $d=1$, so that $x=0$ maps to $x=1$. This is true for chain of $A_0$ matrices of any length and chains starting
with $A_1$ following by a chain of $A_0$ matrices of any length. These two cases give us certain expectation values with $r$ spins on the rhs fixed:
\begin{eqnarray}\label{UUUrinftwospins}
\langle \underbrace {\ldots}_{\infty} \uparrow \underbrace {\ldots}_{n}\underbrace{\uparrow \ldots \uparrow  \ldots \uparrow}_r \rangle_{x=0} 
&=& 
\left ( 1+\frac{2-\lambda}{\lambda^{n+1}}  \right) \frac{1}{2\lambda^{r}}
\end{eqnarray}
and
\begin{eqnarray}\label{DUUUrinftwospins}
\langle \underbrace {\ldots}_{\infty} \uparrow \underbrace {\ldots}_{n}\downarrow\underbrace{\uparrow \ldots \uparrow  \ldots \uparrow}_{r-1} \rangle_{x=0} 
&=& 
\frac{\lambda-1}{2\lambda^{r}}.
\end{eqnarray}
Similarly we get 
\begin{eqnarray}\label{UUUrinftwospinsD}
\langle \underbrace {\ldots}_{\infty} \downarrow \underbrace {\ldots}_{n}\underbrace{\uparrow \ldots \uparrow  \ldots \uparrow}_r \rangle_{x=0} 
&=& 
\left ( 1-\frac{2-\lambda}{\lambda^{n+1}}  \right) \frac{1}{2\lambda^{r}}
\end{eqnarray}
and
\begin{eqnarray}\label{DUUUrinftwospinsD}
\langle \underbrace {\ldots}_{\infty} \downarrow \underbrace {\ldots}_{n}\downarrow\underbrace{\uparrow \ldots \uparrow  \ldots \uparrow}_{r-1} \rangle_{x=0} 
&=& 
\frac{\lambda-1}{2\lambda^{r}}.
\end{eqnarray}
Note that (\ref{DUUUrinftwospins}) and (\ref{DUUUrinftwospinsD}) generalize  (\ref{UDinftwospinssmallr0}) and (\ref{DDinftwospinssmallr0}) in that they  are both independent of $n$ and equal to each other. Thus the restoration of spin symmetry already seen, which holds at any temperature and for any separation $n$, is also valid for any $r$.  This seems very curious and nonÐtrivial.  Some understanding can be gained by considering recent ideas about the mechanism underlying the phase transition \cite{PFK}.  According to this work, the transition is due to the condensation of clusters of spins of exactly the type on the right hand edge in  (\ref{DUUUrinftwospins}) and (\ref{DUUUrinftwospinsD}).  This is consistent with our results here, since it shows that such clusters restore the spin symmetry which is broken by the ``hidden" up spin on the left hand edge of the chain, at least for the particular expectation values investigated.

Note that to calculate any of (\ref{UUinftwospinssmallr0}) - (\ref{DUUUrinftwospinsD}) for $x=1$ would require 
knowing the four values $U_n(1/2)$, $U_n(2)$, $D_n(1/2)$ and $D_n(2)$. Using (\ref{UDrelation2}) accounts for two of these, in addition (\ref{UDrelation1})
removes one more, but one is left with one unknown value. For general $x$, one has four unknown quantities.
 
Finally, summing (\ref{UUUrinftwospins}) and (\ref{UUUrinftwospinsD})  or (\ref{DUUUrinftwospins}) and (\ref{DUUUrinftwospinsD}) gives rise to, respectively,
\begin{eqnarray}\label{UUrinftwospins}
\langle \underbrace {\ldots}_{\infty} \underbrace{\uparrow \ldots \uparrow  \ldots \uparrow}_r \rangle_{x=0} 
&=& 
 \frac{1}{\lambda^{r}}
\end{eqnarray}
and
\begin{eqnarray}\label{DUUrinftwospins}
\langle \underbrace {\ldots}_{\infty} \downarrow\underbrace{\uparrow \ldots \uparrow  \ldots \uparrow}_{r-1} \rangle_{x=0} 
&=& 
\frac{\lambda-1}{\lambda^{r}}.
\end{eqnarray}
As one approaches the phase transition, $\lambda \to 1$, so that (\ref{UUrinftwospins}) goes to $1$ while (\ref{DUUrinftwospins}) approaches $0$.  This suggests that in an infinite chain exactly at the transition, the only state with non-zero probability has all spins up.  This would not be surprising, since the same property holds below the transition (for $\beta > \beta_c$).

\section{Conclusions}
In this paper we have extended our understanding of the statistical mechanical behavior of the Farey spin chains.  Our main tool is a generalization of the ``number theoretic" partition function studied by Knauf \cite{K,C-K,K-o1,C-Kn}.  

By introducing a new parameter, we are able to derive recursion relations on the length of the spin chain (or equivalently, the level of the Farey fractions).  These relations are generalizations of the Lewis three-term equation of number theory \cite{L-Z, LZ}.  Using them and the behavior of the system under spin-flip transformations, we find new results. In particular, we prove a new and simple connection between the Ruelle-Perron-Frobenius transfer operator studied by Prellberg \cite{P-S, P-s, Diss} in a model of dynamical systems with intermittency and our generalized partition function.  This connection implies that all our models have the same free energy (and hence the same thermodynamics), independent of the value of $x$.  It also implies that the correlation length satisfies the prediction of scaling theory. 

In addition, we are able to calculate certain spin expectation values  for both finite and infinite spin chains.  These results are the first such calculations, to our knowledge.  In particular, we find that the expectation value of certain spin clusters and an independent spin at arbitrary distance is independent of the direction of the spin, at all temperatures above the transition. This holds even though the expectation value of the the independent spin by itself does depend on direction.  Thus, in this sense, the spin cluster removes the spin asymmetry of the system.  This behavior appears to be related to the mechanism of the phase transition.

\section{Acknowledgements}
We are grateful to Don Zagier for suggesting the generalized partition function and for help with an early derivation of expectation values.  Useful conversations with Ali  \"{O}zl\"{u}k are also acknowledged.  This work was supported in part by the National Science Foundation Grant No. DMR-0203589.

\end{document}